\documentstyle[12pt]{article}
\newlength{\extraspace}
\newlength{\extraspaces}
\catcode`\@=11
\newcommand{\be}{\begin{equation}
\addtolength{\abovedisplayskip}{\extraspaces}
\addtolength{\belowdisplayskip}{\extraspaces}
\addtolength{\abovedisplayshortskip}{\extraspace}
\addtolength{\belowdisplayshortskip}{\extraspace}}
\newcommand{\ee}{\end{equation}}
\newcommand{\ba}{\begin{eqnarray}
\addtolength{\abovedisplayskip}{\extraspaces}
\addtolength{\belowdisplayskip}{\extraspaces}
\addtolength{\abovedisplayshortskip}{\extraspace}
\addtolength{\belowdisplayshortskip}{\extraspace}}
\newcommand{\ea}{\end{eqnarray}}
\newcommand{\nonu}{\nonumber \\[.5mm]}
\newcommand{\A}{&\!\!\!}

\newcommand{\e}{\, {\rm e}}

\begin{document}
\addtolength{\baselineskip}{.7mm}
\thispagestyle{empty}
\begin{flushright}
TIT/HEP--291 \\
STUPP--95--140 \\
{\tt hep-th/9505132} \\
May, 1995
\end{flushright}
\vspace{2mm}
\begin{center}
{\Large{\bf Dilaton Gravity Coupled to a
Nonlinear Sigma Model in $2 + \epsilon$ Dimensions
}} \\[15mm]
{\sc Shin-ichi Kojima},\footnote{
\tt e-mail: kotori@phys.titech.ac.jp} \hspace{5mm}
{\sc Norisuke Sakai}\footnote{
\tt e-mail: nsakai@phys.titech.ac.jp} \\[3mm]
{\it Department of Physics, Tokyo Institute of Technology \\[2mm]
Oh-okayama, Meguro, Tokyo 152, Japan} \\[4mm]
and \\[4mm]
{\sc Yoshiaki Tanii}\footnote{
\tt e-mail: tanii@th.phy.saitama-u.ac.jp} \\[3mm]
{\it Physics Department,
Saitama University, Urawa, Saitama 338, Japan} \\[15mm]
{\bf Abstract}\\[5mm]
{\parbox{13cm}{\hspace{5mm}
Quantum theory of dilaton gravity coupled to a nonlinear sigma model
with a maximally symmetric target space is studied in $2+\epsilon$
dimensions. The ultraviolet stable fixed point for the curvature of
the nonlinear sigma model demands a new fixed point theory for the
dilaton coupling function. The fixed point of the dilaton coupling
is a saddle point similarly to the previous case of the flat target
space.
}}
\end{center}
\vfill
\newpage
%
%%%%%%%%%%%%%%%%%%%%%%%%%%%%%%%%%%%%%%%%%%%%%%%%%%%%%%%%%%%%%%%%%%%
%
It has been proposed sometime ago \cite{WEI} that the quantum theory
of gravity can consistently be defined by an analytic continuation
from lower dimensions, similarly to the successful application for
nonlinear sigma models \cite{BLS}.
The power-counting arguments necessitate an expansion of the quantum
gravity around two dimensions.
On the other hand, the usual Einstein action becomes a topological
invariant precisely at two dimensions.
The clash between the singular nature of the Einstein action at
two dimensions and the necessity for an expansion around two
dimensions is the fundamental origin of a problem afflicting
the ($2+\epsilon$)-dimensional approach to quantum gravity.
If one takes the conformal gauge, one finds that the propagator
for the Liouville field becomes singular and an oversubtraction
is needed at the one-loop order \cite{KN}.
The singular propagator together with the oversubtraction leads to
the nonrenormalizable divergences at the two-loop order \cite{JJ},
\cite{KKN}.
\par
To avoid the clash between the singular action and the expansion
around two dimensions, it has been proposed to abandon
the general coordinate invariance and to use only
volume-preserving diffeomorphisms as the fundamental principle
\cite{KKNS}.
On the other hand, in order to maintain the general coordinate
invariance,
we have proposed an alternative approach in a preceding paper to
solve the difficulty:
the dilaton gravity can be used to define the higher
dimensional quantum theory of Einstein gravity \cite{KST2}.
We have observed that the dilaton gravity action is equivalent to
the Einstein action with an additional free scalar field, except
in two spacetime dimensions.
Most importantly, the dilaton gravity action is smooth around
two dimensions.
As a result, the Liouville-dilaton propagator becomes nonsingular
even in two dimensions.
Therefore the nonrenormalizable divergences arising from the
singular Liouville propagator should be absent in our
dilaton gravity theory.
We have explicitly studied the case of $N$ free massless scalar fields
as matter fields interacting with the dilaton-gravity system.
We have obtained divergences and beta functions
to one-loop which exhibit a nontrivial fixed point.
The fixed point is ultraviolet stable for the gravitational
coupling constant $G$, if $\epsilon > 0$ and $N < 24$,
although it is not ultraviolet stable for the strength of the
dilaton coupling function \cite{KST2}.
We have also found that the fixed point theory can be transformed to
an action of the usual CGHS type \cite{CGHS}.
The ($2+\epsilon$)-dimensional gravity with or without dilaton have
been applied to other interesting situations
\cite{KST} -- \cite{ELOD}.
\par
The results on the dilaton gravity, however, may depend on the
choice of possible interactions among matter fields.
Therefore it is worthwhile to study
the renormalization group properties of the dilaton gravity further
using models with interacting matter fields.
\par
The purpose of this letter is to study the dilaton gravity coupled
to a nontrivially interacting scalar fields.
For simplicity, we consider a nonlinear sigma model with a maximally
symmetric target space as an interacting matter fields.
We find a nontrivial fixed point which is ultraviolet stable in the
direction of both the gravitational coupling constant $G$ and the
curvature
of the maximally symmetric target space $\alpha'$, but is a saddle
point in the direction of the dilaton coupling function.
It turns out that the dilaton gravity at the fixed point can be
transformed to a free dilaton field embedded into the Einstein
gravity, if fields are redefined with a local Weyl rescaling
which is singular as $\epsilon \rightarrow 0$.
Since the transformation is singular in the limit of two dimensions,
such a transformation is not allowed in our
($2+\epsilon$)-dimensional approach.
Therefore the fixed point theory cannot be regarded as a free
dilaton gravity theory.
The necessity of fine tuning of the dilaton coupling function
is a feature which is common to our previous result on the free
massless scalar fields \cite{KST2}.
On the other hand, the new fixed point of our
nonlinear sigma model with the maximally symmetric target space
corresponds to an ultraviolet stable nontrivial fixed point
($\alpha'\not=0$) for the curvature of the
target space.
This fixed point does not reduce to our previous result
for the free massless scalar fields which corresponds
to an infrared stable trivial fixed point ($\alpha'=0$)
for the curvature.
%Our new fixed point is not connected to the fixed point
%that we have found before, although
Note that both of them require a fine-tuning of the
dilaton coupling function.
\par
Let us consider $N+1$ spinless fields $\phi, X^j$
($j=1, \cdots, N$). We require the power-counting renormalizability
in two-dimensional limit and find the most general
action as a nonlinear sigma model (including the dilaton $\phi$)
coupled to gravity
as
\ba
S \A = \A \mu^\epsilon \int d^d x \sqrt{-g} \biggl[
{1 \over 16\pi G} R^{(d)} L(\phi, X)
- {1 \over 2} g^{\mu\nu} \partial_\mu \phi \partial_\nu \phi
G_{\phi\phi}(\phi, X) \nonu
\A \A - g^{\mu\nu} \partial_\mu \phi \partial_\nu X^j
G_{\phi j}(\phi, X)
- {1 \over 2} g^{\mu\nu} \partial_\mu X^i \partial_\nu X^j
G_{ij}(\phi, X) \biggr],
\label{generalaction}
\ea
which contains four arbitrary functions
$L$, $G_{\phi\phi}$, $G_{\phi j}$, $G_{ij}$ of $\phi$ and $X^i$.
Here, $d = 2+\epsilon$, and $G$ and $\mu$ are the renormalized
gravitational constant and the renormalization scale respectively.
In this paper, we consider a nonlinear sigma model for the
interacting matter fields $X^i$ to form the maximally symmetric
target space in $N$ dimensions as the simplest target space
(of $X^i$) which has a nontrivial geometry.
The maximal symmetry requires the coupling functions to be
\ba
L(\phi,X) \A = \A L(\phi), \qquad
G_{\phi j}(\phi,X)  =  0, \nonu
G_{\phi\phi}(\phi,X) \A = \A \Psi(\phi), \quad
G_{ij}(\phi,X) = {1 \over 2\pi\alpha'} \bar{G}_{ij}(X)
\e^{-2\Phi(\phi)},
\label{dilatonmattermetric}
\ea
where $\alpha'$ is proportional to the curvature of the target
space. The metric of the maximally symmetric space
$\bar{G}_{ij}$ satisfies
\be
\bar{R}_{ijkl} = k\left(\bar{G}_{ik}\bar{G}_{jl} -
\bar{G}_{il}\bar{G}_{jk}\right),
\label{maximalsym}
\ee
where the parameter $k$ takes its value on $\{0, \pm 1\}$.
The symmetry of the space of the matter fields is
$O(N+1)$ for $k=+1$, $O(N,1)$ for $k=-1$ and the $N$-dimensional
Euclidean group for $k=0$ respectively.
The $k=0$ case has been studied in our previous paper \cite{KST2}.
In this paper we study the cases $k=\pm 1$.
To fix the meaning of the curvature of the target space, we
choose $\Phi(0) = 0$.
\par
Similarly to the previous case of the dilaton gravity with free
massless scalar fields \cite{KST2},
we have two kinds of freedom to redefine the fields.
The first one is
the local Weyl rescaling of the metric
%
%\be
$
g_{\mu\nu} \rightarrow \e^{-2\Lambda(\phi)} g_{\mu\nu}
$
%\label{localweylrescale}
%\ee
%
with a function $\Lambda$ of the dilaton $\phi$.
This can be used to
fix one of the functions $L, \Psi$, or $ \Phi$.
Since the function $\Psi$ changes by terms of order
$\epsilon^0$ by the local Weyl rescaling, whereas
$L, \Phi$ change only by terms of order $\epsilon$,
we choose to fix the function $\Psi$ by means of
the local Weyl rescaling.
We can find the local Weyl rescaling $\Lambda$ which transforms a
generic model to the model with $\Psi(\phi)=0$, and is finite as
we let $\epsilon \rightarrow 0$.
Let us note that this choice fixes only $\Lambda'(\phi)$, namely
the nonzero modes of $\Lambda(\phi)$.
The second field redefinition is
an arbitrary field redefinition of the dilaton $\phi$
with a function $f$ of the dilaton
%
%\be
$\phi \rightarrow f(\phi)$.
%\label{dilatonredef}
%\ee
%
We can use it to fix the form of the function $L(\phi)$, which we
choose $L(\phi)=\exp(-2\phi)$ as the standard form.
This fixes only nonzero modes of $f(\phi)$.
The zero modes of $\Lambda(\phi)$ and $f(\phi)$ are used
to fix coefficients of two reference operators, analogously to
the case of free massless scalar fields \cite{KST2}.
With these choices of $\Lambda(\phi)$ and $f(\phi)$, we obtain the
following standard form of the action of dilaton gravity coupled
to the nonlinear sigma model
\be
S = \mu^\epsilon \int d^d x \sqrt{-g} \biggl[ {1 \over 16\pi G}
R^{(d)} \e^{-2\phi} - {1 \over 4\pi\alpha'} g^{\mu\nu}
\partial_\mu X^i \partial_\nu X^j
\bar{G}_{ij}(X) \e^{-2\Phi(\phi)} \biggr],
  \label{maximalsymaction}
\ee
where $\Phi(0) = 0$.
\par
We use the background field method \cite{DEWITT}
to compute one-loop divergences.
Let us introduce the background metric $\hat g_{\mu\nu}$ and
decompose the metric $g_{\mu\nu}$ into the traceless
field $h_{\mu\nu}$ and the Liouville field $\rho$
\be
g_{\mu\nu} = \tilde g_{\mu\nu} \e^{-2\rho}
= \hat g_{\mu\lambda} (\e^{\kappa h})^\lambda{}_\nu \e^{-2\rho},
\label{metricdec}
\ee
where $\kappa^2 = {16 \pi G \over \mu^\epsilon}$.
The action (\ref{maximalsymaction}) becomes
\ba
S \A = \A \mu^\epsilon \int d^d x \sqrt{-\tilde g}
\e^{-\epsilon\rho} \biggl[ {1 \over 16 \pi G}
\biggl(\tilde R^{(d)} \e^{-2\phi}
+ \epsilon(\epsilon+1) \tilde g^{\mu\nu} \partial_\mu \rho
\partial_\nu \rho \e^{-2\phi} \nonu
\A \A + 4 (\epsilon+1) \tilde g^{\mu\nu} \partial_\mu \rho
\partial_\nu \phi \e^{-2\phi} \biggr)
- {1 \over 4 \pi \alpha'} \tilde g^{\mu\nu} \partial_\mu X^i
\partial_\nu X^j \bar{G}_{ij}(X) \e^{-2\Phi(\phi)} \biggr].
\label{dilatonaction}
\ea
At this point we emphasize the following point for the nonsingular
nature of our dilaton gravity theory.
We can readily read off the propagator for the Liouville-dilaton
system from the above action and find that they have smooth
two-dimensional limit $\epsilon \rightarrow 0$.
More generally, even if we allow the most general coupling between
dilaton and other matter fields including $G_{\phi j}(\phi, X)\not=0$ in
Eq.\ (\ref{dilatonmattermetric}),
we find that the propagator of all fields are nonsingular
in the limit of two dimensions provided the matter kinetic terms
are nonsingular ${\rm det} G_{ij}\not=0$.
Therefore the nonsingular propagator and the smoothness of the
dilaton gravity theory is a generic feature irrespective of
the details of possible couplings among matter and dilaton fields.
This point is crucial \cite{KST2} to eliminate nonrenormalizable
divergences observed in Refs.\ \cite{JJ}, \cite{KKN}.
\par
Let us make use of our previous result for one-loop calculation
for a general action \cite{KST2} involving the Liouville field
together with the dilaton and matter fields.
Combining all the scalar fields and the Liouville field into
$Y^I = (\rho, \phi, X^i)$, where the index runs
$I =(\rho, \phi, i)$, we obtain a kind of nonlinear sigma model
using the metric $\tilde{g}_{\mu\nu} = \hat{g}_{\mu\lambda}
(\e^{\kappa h})^{\lambda}{}_{\nu}$
with the Liouville field $\rho$ removed
\ba
S = {\mu^\epsilon \over 16\pi G}
\int d^d x \sqrt{-\tilde g} \left[ \tilde R^{(d)} L(Y)
- {1 \over 2} \tilde g^{\mu\nu} \partial_\mu Y^I
\partial_\nu Y^J G_{IJ}(Y) \right],
  \label{actiony}
\ea
where the curvature for the
metric $\tilde g_{\mu\nu}$ is denoted as $\tilde R^{(d)}$.
The functions $L(\phi)$ and $G_{IJ}$ in our specific model can be read
off from Eq.\ (\ref{dilatonaction}).
The gauge fixing term and the ghost action are given in terms of
the gauge fixing function $F_{\alpha}$ \cite{KST2} as
\ba
\A \A S_{\rm GF+FP} = \int d^d x \, \delta_{\rm B} \bigl(
- i b^\alpha F_\alpha \bigr), \nonu
\A \A F_\alpha =
\sqrt{-\hat g} \hat L \left[ \hat D^\beta h_{\beta\alpha}
- {1 \over \kappa} \partial_\alpha \left( {L(Y) \over \hat L}
\right) + {1 \over 2} B_\alpha \right],
\label{gffp}
\ea
where $\delta_{\rm B}$ is the BRST transformation and the fields
with a hat are background.
The one-loop divergences in the effective action of the general
model (\ref{actiony}) with the gauge fixing term and the ghost
action (\ref{gffp}) is given by \cite{KST2}
\be
\Gamma_{\rm div}
= \int d^d x \sqrt{-\hat g} \biggl[
{24-N \over 24\pi\epsilon} \hat R^{(d)} - {1 \over 4\pi\epsilon}
\hat g^{\mu\nu} \partial_\mu \hat Y^I \partial_\nu \hat Y^J
\left( \hat R_{IJ} + \partial_I \ln \hat L
\partial_J \ln \hat L \right) \biggr].
  \label{generaldivergence}
\ee
Let us apply this result to our model (\ref{maximalsymaction}).
By computing the curvature in the $Y^I=(\rho, \phi, X^i)$ space
from the metric in Eq.\ (\ref{dilatonaction}),
we obtain the one-loop divergence %and the counter terms
of the dilaton gravity coupled to
the maximally symmetric nonlinear sigma model.
The counter terms can be summarized with
three types of coefficients $A, B, C$
\ba
S_{\rm counter} \A = \A - \mu^\epsilon \int d^d x \sqrt{-g}
\biggl[ R^{(d)} A(\phi)
+ g^{\mu\nu} \partial_\mu \phi \partial_\nu \phi B(\phi) \nonu
\A \A - {1 \over 2} g^{\mu\nu} \partial_\mu X^i \partial_\nu X^i
\bar{G}_{ij}(X) C(\phi) \biggr].
\label{dilcounter}
\ea
We find that these coefficients are given at one-loop order by
\ba
A(\phi) \A = \A {24-N \over 24\pi\epsilon}, \nonu
B(\phi) \A = \A - {1 \over \pi\epsilon} + {N \over 4\pi\epsilon}
\left[ \left( \Phi'(\phi) \right)^2 - \Phi''(\phi)
- 2 \Phi'(\phi) \right], \nonu
C(\phi) \A = \A {N-1 \over 2 \pi \epsilon} \, k.
\ea
\par
Let us renormalize the dilaton gravity.
The action including counter terms is given in terms of bare
quantities denoted by the suffix $0$ as
\ba
S_0 \A = \A S + S_{\rm counter} \label{bareaction} \\
\A = \A \int d^d x \sqrt{-g_0} \left[ {1 \over 16\pi G_0}
R^{(d)}_0 \e^{-2\phi_0} - {1 \over 4 \pi \alpha_0'} g_0^{\mu\nu}
\partial_\mu X^i \partial_\nu X^j \bar{G}_{ij}(X)
\e^{-2\Phi_0(\phi_0)} \right]. \nonumber
\ea
By defining the renormalized quantities as
\ba
\Phi_0(\phi_0) \A = \A \Phi(\phi) + F(\phi) \qquad (F(0) = 0), \nonu
g_{0\;\mu\nu} \A = \A g_{\mu\nu} \e^{-2\Lambda(\phi)} \qquad
(\Lambda(0) = 0), \nonu
\phi_0 \A = \A \phi + f(\phi) \qquad (f(0) = 0),
\label{barequantities}
\ea
the renormalization of the one-loop divergence can be achieved
by requiring
\ba
\A \A {1 \over 16\pi G_0}
\e^{- 2\phi - \epsilon\Lambda(\phi) - 2 f(\phi)}
= \mu^\epsilon \biggl[ {1 \over 16\pi G} \e^{-2\phi}
- A(\phi) \biggr], \nonu
\A \A {\epsilon+1 \over 16\pi G_0} \Bigl(
4 \Lambda' + \epsilon (\Lambda')^2 + 4 f' \Lambda' \Bigr)(\phi)
\e^{- 2\phi - \epsilon\Lambda(\phi) - 2 f(\phi)}
= - \mu^\epsilon B(\phi), \nonu
\A \A {1 \over 2\pi\alpha_0'}
\e^{- 2\Phi(\phi)-\epsilon\Lambda(\phi)-2F(\phi)}
= \mu^\epsilon \biggl[ {1 \over 2\pi\alpha'} \e^{- 2\Phi(\phi)}
- C(\phi)\biggr].
\ea
Using the fact that $A(\phi)$, $C(\phi)$ are constant,
$\Lambda, f, F = O(G, \alpha')$ at one-loop order,
and neglecting higher order
terms in $G$, $\alpha'$ we find the solution of these equations
and obtain the relation between the bare and renormalized
quantities as
\ba
{1 \over G_0} \A = \A \mu^\epsilon \biggl( {1 \over G}
- 16\pi A \biggr), \qquad
{1 \over \alpha_0'}  =  \mu^{\epsilon}
\biggl( {1 \over \alpha'} - 2 \pi C \biggr), \nonu
\rho_0 \A = \A \rho - {4\pi G \over \epsilon+1}
\int_0^\phi d \phi' \e^{2\phi'} B(\phi'), \nonu
\phi_0 \A = \A \phi + 8\pi A G \left( \e^{2\phi} - 1 \right)
+ {2\pi \epsilon G \over \epsilon+1}
\int_0^\phi d \phi' \e^{2\phi'} B(\phi'), \nonu
\Phi_0(\phi_0) \A = \A \Phi(\phi)
+ \pi \alpha' C \left( \e^{2\Phi(\phi)} - 1 \right)
+ {2\pi \epsilon G \over \epsilon+1}
\int_0^\phi d \phi' \e^{2\phi'} B(\phi') \nonu
\A = \A \Phi(\phi_0) - 8\pi G A \Phi'(\phi_0) (\e^{2\phi_0} - 1)
+ \pi\alpha' C \left( \e^{2\Phi(\phi_0)} - 1 \right)  \nonu
\A \A - {2\pi\epsilon G \over \epsilon + 1}(\Phi'(\phi_0) - 1)
\int^{\phi_0}_{0} d\phi' \e^{2\phi'} B(\phi').
\ea
We find that the beta functions $\beta$ and the anomalous
dimensions $\gamma$ are functions of $\phi$ in general
\ba
\beta_G \A \equiv \A \mu {\partial G \over \partial \mu}
= \epsilon G - 16 \pi \epsilon A G^2,
%\nonu
\qquad
\beta_{\alpha'}
%\A
\equiv
%\A
\mu {\partial \alpha' \over \partial \mu}
= \epsilon \alpha' - 2 \pi \epsilon C \alpha'^2, \nonu
\beta_\Phi(\phi_0)
\A  \equiv  \A
\mu {\partial \Phi(\phi_0) \over \partial \mu} \nonu
\A   =  \A
8\pi \epsilon A G \Phi'(\phi_0) \left( \e^{2\phi_0} - 1 \right)
- \pi \epsilon \alpha' C \left( \e^{2\Phi(\phi_0)} - 1\right)
\nonu
\A \A + {2\pi \epsilon^2 G \over \epsilon+1}
\left( \Phi'(\phi_0) - 1 \right)
\int_0^{\phi_0} d \phi' \e^{2\phi'} B(\phi'), \nonu
\gamma_\rho
\A  \equiv \A  \mu {\partial \rho \over \partial \mu}
= {4\pi \epsilon G \over \epsilon+1}
\int_0^\phi d \phi' \e^{2\phi'} B(\phi'), \nonu
\gamma_\phi \A  \equiv \A  \mu {\partial \phi \over \partial \mu}
= - 8\pi \epsilon A G \left( \e^{2\phi} - 1 \right)
- {2\pi \epsilon^2 G \over \epsilon+1}
\int_0^\phi d \phi' \e^{2\phi'} B(\phi').
\label{betafunction}
\ea
\par
The beta function $\beta_G$ is identical to that of the Einstein
gravity \cite{KN} as was reported in \cite{KST2}.
For $N < 24$ and $\epsilon > 0$, $G = 0$ is an infrared stable fixed
point and $G = G_\ast$ is an ultraviolet stable fixed point, where
\be
G_{\ast} = {3\epsilon \over 2(24 - N)}, \qquad
\beta_G (G_{\ast}) = 0, \qquad
\beta'_G (G_{\ast}) < 0.
\label{gravfp}
\ee
The beta function $\beta_{\alpha'}$  coincides with that of a
nonlinear $\sigma$-model with a maximally symmetric target space at
one-loop order. For $\epsilon > 0$, the trivial fixed point
$\alpha' = 0$ is infrared stable and the nontrivial fixed point
$\alpha' = \alpha'_{\ast}$ is
ultraviolet stable, irrespective of the sign of the
target space curvature $k = \pm 1$
\be
\alpha'_{\ast} = {\epsilon \over (N - 1) k}, \qquad
\beta_{\alpha'}(\alpha'_{\ast}) = 0, \qquad
\beta'_{\alpha'}(\alpha'_{\ast}) < 0.
\label{nlsigmafp}
\ee
The fixed point condition $\beta_\Phi(\phi)= 0$ for the dilaton
coupling function $\Phi(\phi)$ is given by a functional equation
\ba
\A \A {24-N \over 3}G ({\rm e}^{2\phi}-1)\Phi'(\phi)
- {(N-1)k \over 2} \alpha' \left( \e^{2\Phi(\phi)} - 1\right)
\label{fixedpointphi} \\
\A \A \qquad + {2\pi \epsilon^2 G \over \epsilon+1}
\left( \Phi'(\phi) - 1 \right)
\int_0^{\phi} d \phi' \e^{2\phi'}
\left(-{1 \over \pi\epsilon} + {N \over 4\pi\epsilon}\left[(\Phi')^2
-\Phi''-2\Phi'\right](\phi') \right) = 0. \nonumber
\ea
An appropriate differentiation leads to
a nonlinear differential equation
\ba
\A \A {(24-N)G \over 3}\left[\left({\rm e}^{2\phi}-1\right)\Phi''
-2{\rm e}^{2\phi}\Phi'\left(\Phi'-1\right)\right] \nonu
\A \A \qquad -
{(N-1)k\alpha' \over 2}\left[\left({\rm e}^{2\Phi}-1\right)\Phi''
-2{\rm e}^{2\Phi}\Phi'\left(\Phi'-1\right)\right] \nonu
\A \A \qquad + {2\pi \epsilon^2 G \over \epsilon+1}
\left( \Phi' - 1 \right)^2  \e^{2\phi}
\left({1 \over \pi\epsilon}-{N \over 4\pi\epsilon}\left[(\Phi')^2
-\Phi''-2\Phi'\right]\right) = 0.
\label{nonlineareq}
\ea
This equation simplifies at the fixed point for $G=G_*$
and $\alpha'=\alpha'_*$. We find immediately that there is a
simple solution
%for the nonlinear equation
%
\be
\Phi_\ast (\phi) =\phi,
\label{dilatonfp}
\ee
although we are unable to find general solutions for this nonlinear
equation. In particular, we find no consistent solution with
$\Phi(\phi)=\lambda \phi$, $\lambda = O(\epsilon)$ which is similar
to the solution for our previous case of free massless scalar
fields for matter \cite{KST2}.
Since the above nonlinear equation is obtained by differentiating
the fixed point condition, it is a necessary but not
a sufficient condition for the fixed point.
Therefore we have checked that the above solution (\ref{dilatonfp})
really satisfies
the fixed point condition $\beta_\Phi(\phi)=0$ in
Eq.\ (\ref{fixedpointphi}).
\par
Let us study the stability of this fixed point (\ref{gravfp}),
(\ref{nlsigmafp}) and (\ref{dilatonfp}).\footnote{
    We would like to correct
    the stability argument in Ref.\ \cite{KST2}.
    The correct form of the linearized beta functions Eq.\ (5.16)
    in Ref.\ \cite{KST2} should be
    $\beta_G = -\epsilon\delta G,\
    \beta_{\Phi} = - {1 \over 2}\epsilon
    (1 - \e^{2\phi}) {d \over d\phi}\delta\Phi + O(\epsilon^2)$.
    Namely these should be diagonal from the beginning,
    so that the Eqs.\ (5.18), (5.20), (5.21) and (5.22) are valid
    for $\Phi$ instead of $\tilde\Phi$.
    Therefore we obtain the same conclusions on the stability and we
    need not introduce $\tilde\Phi$ defined in
    Eq.\ (5.17). Then the fine-tuning of $\Phi$ in Eq.\ (5.24) is
    simply $\Phi(\phi) = \lambda^\ast \phi$.}
We expand the beta functions around the fixed point
\be
G = G_{\ast} + \delta G, \qquad \alpha'
= \alpha'_{\ast} + \delta \alpha',
\qquad \Phi(\phi) = \phi + \delta \Phi(\phi),
\ee
assuming the fluctuations $\delta G$, $\delta \alpha'$ and
$\delta\Phi$ to be small. By examining the beta functions around
the fixed point to first order in the fluctuations we find
\ba
\beta_G \A = \A - \epsilon \delta G, \qquad
\beta_{\alpha'} = - \epsilon \delta \alpha', \nonu
\beta_{\Phi} \A = \A \left( \e^{2\phi} - 1 \right)
\biggl[ {24 - N \over 3} \delta G - {(N - 1) k \over 2}
\delta \alpha' + {\epsilon \over 2} \delta \Phi' \biggr]
- \epsilon \e^{2\phi} \delta \Phi.
\ea
We diagonalize these equations by introducing the following
quantities
\ba
\tilde\Phi(\phi) \A \equiv \A \Phi(\phi) + X(\phi) G
+ Y(\phi) \alpha', \nonu
X(\phi) \A \equiv \A - {1 \over 2 G_{\ast}}
\left( 1 - \e^{2\phi} \right), \qquad
Y(\phi) \equiv {1 \over 2\alpha'_{\ast}}
\left( 1 - \e^{2\phi} \right).
\ea
Then we consider the beta function for $\tilde\Phi$ and obtain
\be
\beta_{\tilde\Phi} = - \biggl[ {\epsilon \over 2}
\left( 1 - \e^{2\phi} \right) {d \over d \phi}
+ \epsilon \e^{2\phi} \biggr] \delta\tilde\Phi.
  \label{diagdeviation}
\ee
If we define a variable $\psi$ to describe the weak coupling
regions ($0<{\rm e}^{\phi}<1$) of
the loop expansion parameter ${\rm e}^{\phi}$ \cite{KST2},
\be
\psi = {1 \over 2} \ln (\e^{-2\phi} - 1), \qquad
\left\{
\begin{array}{l}
\psi \rightarrow + \infty \Longleftrightarrow
\phi \rightarrow - \infty, \\
\psi \rightarrow - \infty \Longleftrightarrow
\phi \rightarrow 0,
\end{array}
\right.
\label{defx}
\ee
Eq.\ (\ref{diagdeviation}) becomes
\be
\beta_{\tilde\Phi} = \biggl[ {\epsilon \over 2} {d \over d \psi}
- {\epsilon \over \e^{2\psi} + 1} \biggr] \delta\tilde\Phi.
\ee
We find eigenfunctions of the differential operator
%on the right hand side
with eigenvalues $\Lambda$ as
\be
\delta\tilde\Phi
= {2 \over \e^{2\psi} + 1}
\e^{\left( 2 + {2 \over \epsilon}\Lambda \right) \psi}
= 2 \e^{2\phi} \left( \e^{-2\phi} - 1
\right)^{(1+ {\Lambda \over \epsilon})},
  \quad
%label{eigenfunction}
%\ee
%
%
%\be
\beta_{\tilde\Phi} = \Lambda \delta \tilde\Phi.
\ee
The condition $\delta\tilde\Phi(\phi = 0) = 0$ requires
$\Lambda > - \epsilon$.
Therefore the fixed point is a saddle point:
ultraviolet stable for direction $- \epsilon < \Lambda < 0$,
and unstable for $\Lambda > 0$.
\par
We can consider a renormalized theory with the gravitational coupling
constant and the target space curvature near the fixed point $G_\ast$,
$\alpha'_\ast$, as long as we fine tune the dilaton coupling function
$\tilde\Phi$ to be
precisely at the fixed point $\delta\tilde\Phi = 0$
\ba
\Phi(\phi) \A = \A \phi - X(\phi)(G - G_{\ast})
- Y(\phi)(\alpha' - \alpha'_{\ast}) \nonu
\A = \A \phi + {1 \over 2} \left( {G \over G_{\ast}}
- {\alpha' \over \alpha'_{\ast}} \right)
\left( 1 - \e^{2\phi} \right).
\ea
After fine-tuning of the dilaton coupling function,  we find the
dilaton gravity theory around the fixed points for the
gravitational constant $G$ and the curvature of the target space
$\alpha'$
\ba
S \A = \A \mu^{\epsilon} \int d^d x \sqrt{-g} \e^{-2\phi}
\biggl[ {1 \over 16 \pi G} R^{(d)} \nonu
\A \A - {1 \over 4 \pi \alpha'} \e^{-\left( {G \over G_{\ast}}
- {\alpha' \over \alpha'_{\ast}} \right)(1 - \e^{2\phi})} g^{\mu\nu}
\partial_\mu X^i \partial_\nu X^j \bar{G}_{ij}(X) \biggr].
\label{actionfinetuned}
\ea
\par
Let us examine the meaning of the fixed point theory for
the dilaton gravity coupled to
the maximally symmetric nonlinear sigma model.
If we tentatively allow a singular Weyl rescaling
$g_{\mu\nu} \rightarrow g_{\mu\nu} \e^{ {4 \over \epsilon} \phi}$,
we obtain
\ba
S \A = \A \mu^{\epsilon} \int d^d x \sqrt{-g}
\biggl[ {1 \over 16 \pi G_{\ast}} \left( R^{(d)}
- {4(1 + \epsilon) \over \epsilon}
g^{\mu\nu} \partial_\mu \phi \partial_\nu \phi \right) \nonu
\A \A - {1 \over 4 \pi \alpha'_{\ast}} g^{\mu\nu}
\partial_\mu X^i \partial_\nu X^j \bar{G}_{ij}(X) \biggr].
        \label{actionfp}
\ea
This is the Einstein gravity coupled to a free scalar $\phi$ and to
scalar fields $X^i$ which form a maximally symmetric target space.
The dilaton field can be put into a free scalar field embedded into
the Einstein gravity as long as we stay away from two dimensions.
Let us emphasize, however, that the ($2+\epsilon$)-dimensional
approach does not permit such a singular local Weyl rescaling and
the fixed point theory cannot be regarded as a truly free
dilaton theory.
\par
We finally clarify a possible relation between the model studied
in this paper and the model in our previous paper \cite{KST2}.
At the tree level, the theory of the dilaton gravity coupled
to free scalar
fields can be reproduced from the dilaton gravity coupled to
matter fields with a maximally symmetric target space by
taking the limit $\alpha' \rightarrow 0$. This can be easily
understood from the fact that the parameter $\alpha'$ is
the curvature of the target space. The fixed point theory,
however, is not connected to that of the dilaton
gravity coupled to free matter fields.
Let us compare the fixed point corresponding to the
nonlinear sigma model
\be
G_\ast = {3 \epsilon \over 2(24 - N)}, \qquad
{\alpha'}_\ast = {\epsilon \over (N-1)k}, \qquad
\Phi_\ast (\phi) = \phi,
\ee
and the fixed point corresponding to the previous model
with free massless scalar matter fields \cite{KST2}
\be
G_\ast = {3 \epsilon \over 2(24 - N)}, \qquad
{\alpha'}_\ast = 0, \qquad
\Phi_\ast (\phi) = - {3\epsilon \over 24 - N} \phi.
\ee
Actually these are two different fixed points in the coupling
constant space. In the direction of $\alpha'$, the former fixed
point is ultraviolet stable. On the other hand, the latter is
ultraviolet unstable and a fine-tuning is necessary. This fact is
the origin of the different forms of the dilaton-matter coupling
function $\Phi_\ast(\phi)$. Even using the freedom of field
redefinitions, we cannot bring our new fixed point theory to the
previous one.
\par
%
%\vspace{5mm}
%
%%%%%%%  Acknowledgement  %%%%%%%%%%%%%%%%%%%%%%%%%%%%%%%%%%%
%
This work is supported in part by Grant-in-Aid for
Scientific Research (S.K.) and (No.05640334) (N.S.)
from the Ministry of Education, Science and Culture.
%
%%%%%%%  References  %%%%%%%%%%%%%%%%%%%%%%%%%%%%%%%%%%%%%%%
%\vspace{5mm}
%\newpage
%
\newcommand{\NP}[1]{{\it Nucl.\ Phys.\ }{\bf #1}}
\newcommand{\PL}[1]{{\it Phys.\ Lett.\ }{\bf #1}}
\newcommand{\CMP}[1]{{\it Commun.\ Math.\ Phys.\ }{\bf #1}}
\newcommand{\MPL}[1]{{\it Mod.\ Phys.\ Lett.\ }{\bf #1}}
\newcommand{\IJMP}[1]{{\it Int.\ J. Mod.\ Phys.\ }{\bf #1}}
\newcommand{\PR}[1]{{\it Phys.\ Rev.\ }{\bf #1}}
\newcommand{\PRL}[1]{{\it Phys.\ Rev.\ Lett.\ }{\bf #1}}
\newcommand{\PTP}[1]{{\it Prog.\ Theor.\ Phys.\ }{\bf #1}}
\newcommand{\PTPS}[1]{{\it Prog.\ Theor.\ Phys.\ Suppl.\ }{\bf #1}}
\newcommand{\AP}[1]{{\it Ann.\ Phys.\ }{\bf #1}}

\begin{thebibliography}{100}
%
\bibitem{WEI} S. Weinberg, in General Relativity, an Einstein
        Centenary Survey, eds.\ S.W. Hawking and W. Israel
        (Cambridge University Press, 1979) p.\ 790;
        R. Gastmans, R. Kallosh and C. Truffin, \NP{B133}
        (1978) 417;
        S.M. Christensen and M.J. Duff, \PL{B79} (1978) 213.
%
\bibitem{BLS} W.A. Bardeen, B.W. Lee and R.E. Shrock,
        \PR{D14} (1976) 985;
        E. Br\'ezin, J. Zinn-Justin and J.C. Guillou,
        \PR{D14} (1976) 2615.
\bibitem{KN} H. Kawai and M. Ninomiya, \NP{B336} (1990) 115.
\bibitem{JJ} I. Jack and D.R.T. Jones \NP{B358} (1991) 695.
\bibitem{KKN} H. Kawai, Y. Kitazawa and M. Ninomiya,
        \NP{B393} (1993) 280.
\bibitem{KKNS} H. Kawai, Y. Kitazawa and M. Ninomiya,
        \NP{B404} (1993) 684;
        T. Aida, Y. Kitazawa, H. Kawai and M. Ninomiya,
        \NP{B427} (1994) 158;
        T. Aida, Y. Kitazawa, J. Nishimura and A. Tsuchiya,
        preprint TIT/HEP--275, KEK--TH--423, UT-Komaba/94--22,
        hep-th/9501056.
\bibitem{KST2} S. Kojima, N. Sakai and Y. Tanii,
        \NP{B426} (1994) 223.
%
\bibitem{CGHS} C.G. Callan, S.B. Giddings, J. Harvey and
        A. Strominger, \PR{D45} (1992) R1005.
\bibitem{KST} S. Kojima, N. Sakai and Y. Tanii,
        \PL{B322} (1994) 59;
        \IJMP{A31} (1994) 5415.
\bibitem{NTT} J. Nishimura, S. Tamura and A. Tsuchiya,
        \IJMP{A10} (1995) 859; \MPL{A9} (1994) 3565.
\bibitem{AIKI} T. Aida and Y. Kitazawa, preprint
        TIT/HEP--288, hep-th/9504075.
\bibitem{ELOD} E. Elizalde and S. Odintsov, \PL{B313} (1993) 347;
        \PL{B347} (1995) 211; preprint CEAB 95/4--10,
        hep-th/9504093.
\bibitem{DEWITT} B.S. De Witt, \PR{\bf 162} (1967) 1195, 1239;
        L.F. Abbott, \NP{B185} (1981) 189;
        G. 't Hooft and M. Veltman,
        {\it Ann.\ Inst.\ Henri Poincar\' e} {\bf 20} (1974) 69.
%
\end{thebibliography}
\end{document}